\documentclass{svmult}

\usepackage{graphicx}    % standard LaTeX graphics tool
\usepackage{multicol}    % used for the two-column index
\usepackage{amsmath}
\begin{document}

\title*{The overlap Dirac operator as a continued fraction}
%\titlerunning{The overlap operator as a continued fraction}
\author{Urs Wenger\inst{1,}\inst{2}}
\institute{Theoretical Physics, Oxford University, 1 Keble Road, Oxford OX1 3NP, UK 
%\texttt{wenger@thphys.ox.ac.uk}
\and NIC/DESY Zeuthen, Platanenallee 6, D--15738 Zeuthen, Germany \texttt{urs.wenger@desy.de}}

\maketitle

%%%%%%%%%%%%%%%%%%%%%%%%%%%%%%%%%%%%%%%%%%%%%%%%%%%%%%%%%%%%%%%%%%%%%%%%%

We use a continued fraction expansion of the sign--function in order to
obtain a five dimensional formulation of the overlap lattice Dirac
operator. Within this formulation the inverse of the overlap operator
can be calculated by a single Krylov space method and nested
conjugate gradient procedures are avoided. We point out that the five
dimensional linear system can be made well conditioned using
equivalence transformations on the continued fractions.

%%%%%%%%%%%%%%%%%%%%%%%%%%%%%%%%%%%%%%%%%%%%%%%%%%%%%%%%%%%%%%%%%%%%%%%%%
\section{Introduction}
\label{sec:intro}
%%%%%%%%%%%%%%%%%%%%%%%%%%%%%%%%%%%%%%%%%%%%%%%%%%%%%%%%%%%%%%%%%%%%%%%%%

Let us start with noting the overlap Dirac operator $D$
describing chirally symmetric fermions on the lattice \cite{Neuberger:1998fp}, 
\begin{equation}
D = \frac{1}{2} \Big(1 + \gamma_5 \text{sign}\big(H(-m)\big)\Big),
\end{equation}\label{eq:overlap Dirac operator}
where $H(-m)$ is a hermitian Dirac operator with a negative mass
parameter $-m$ of the order of the cut-off of the lattice theory, $m
\sim O(a)$. A bare quark mass $\mu$ is most conveniently introduced as
$D(\mu) = (1-\mu) D + \mu$.  In order to calculate efficiently the
sign--function in eq.(\ref{eq:overlap Dirac operator}) one can use a
rational approximation $\text{sign}(x) \simeq R_{n,m}(x)$ where
$R_{n,m}$ is a (nondegenerate and irreducible) rational function with
algebraic polynomials of order $n$ and $m$ as numerator and
denominator, respectively. Rational approximations usually converge
much faster with their degree than polynomial approximations, but of
course for our specific application it might still be much more
expensive to apply the low degree denominator of the rational function
than to apply a high order polynomial. However, noting that a rational
function can be written as a partial fraction by matching poles and
residues, i.e., $R_{n,m}(x) \sim x \sum_{k} c_k/(x^2+q_k)$, one can
use a multi--shift linear system solver where the convergence is
governed by the smallest of the shifts $q_k$. The overall cost is
therefore roughly equivalent to one inversion of $H^2$.

In order to do physics we need to compute inverses of the overlap
operator $D(\mu)$ to obtain propagators, fermionic forces for
Hybrid Monte Carlo, etc. If we consider the multi--shift linear system
above, we realise that the inversion of $D(\mu)$ leads to a
two--level nested linear system solution, which is rather cumbersome and
forbidding. It is well known how this can be avoided \cite{Ne99,Bo02}: by
introducing a continued fraction representation of the rational
approximation and auxiliary fields, the non--linear system $D(\mu) \psi
= \chi$ can be unfolded into a set of systems linear in $H$. The
auxiliary fields can be interpreted as fields living in a fictitious
fifth dimension and in this way the nested 4D Krylov space problem
reduces to finding a solution in a single 5D Krylov space.  One can
also regard the auxiliary fields as additional fermion flavours which,
when integrated out, generate an effective Dirac operator equivalent
to $D(\mu)$.

%%%%%%%%%%%%%%%%%%%%%%%%%%%%%%%%%%%%%%%%%%%%%%%%%%%%%%%%%%%%%%%%%%%%%%%%%
\section{Matrix representation of the $\text{sign}$--function}
\label{sec:matrix representation}
%%%%%%%%%%%%%%%%%%%%%%%%%%%%%%%%%%%%%%%%%%%%%%%%%%%%%%%%%%%%%%%%%%%%%%%%%

Consider a rational approximation to the $\text{sign}$--function and
expand it as a continued fraction\footnote{Note that since polynomial
approximations can also be expanded into continued fractions all our
considerations apply to them as well.},
\begin{equation} 
R_{2n+1,2n}(x) =
\alpha_0 x + \cfrac{\alpha_1}{x+ \cfrac{\cdots}{\cdots +
	\cfrac{\alpha_{2n}}{x}}} \, .  
\label{eq:rational function as a continued fraction}
\end{equation}
If we rewrite the linear system
\begin{equation} 
	\left(\alpha_0 x + \frac{\alpha_1}{x+ \frac{\cdots}{\cdots +
	\frac{\alpha_{2n}}{x}}} \right) \psi = \chi	
\end{equation}
using appropriate auxiliary fields $\phi_1, \phi_2, \ldots, \phi_{2n}$
we obtain the system
\begin{equation} 
\left( \begin{array}{cccccc} 
      \alpha_0 x & \sqrt{\alpha_1}  &   &  & & \\ 
  \sqrt{\alpha_1}& -x               & \sqrt{\alpha_2}  & & &  \\ 
                 &  \sqrt{\alpha_2} & x  & & &  \\
		 &                  &    & \ddots & & \\ 
                 &                  & & & -x & \sqrt{\alpha_{2n}}\\
	         &                  & & & \sqrt{\alpha_{2n}} & x \end{array} \right) \left
	( \begin{array}{c} \psi \\ \phi_1\\ \phi_2 \\ \vdots \\ \phi_{2n-1} \\ \phi_{2n}
	\end{array} \right) = \left( \begin{array}{c} \chi \\ 0 \\ 0 \\
	\vdots \\ 0 \\ 0 \end{array} \right). 
\end{equation}
By performing explicitly a Schur decomposition of the matrix it is
easy to see that the inverse of eq.(\ref{eq:rational function as a
continued fraction}) is just the $(1,1)$--component of the Schur
complement.

The rational function can also be mapped into a so--called simple
continued fraction
\begin{equation}
R_{2n+1,2n}(x) = \beta_0 x + \cfrac{1}{\beta_1 x+ \cfrac{\cdots}{\cdots +
\cfrac{1}{\beta_{2n} x}}} 
\end{equation}
which leads to a different matrix
\begin{equation}
\left( \begin{array}{cccccc}
	\beta_0 x      & 1 &                 &  &  & \\
	1 & - \beta_1 x              & 1 &  &  & \\
	                & 1 & \beta_2 x               &  &  & \\
                        &                 &           & \ddots &  &\\
                        &                 &           &      &  
	- \beta_{2n-1 } x & 1 \\ 
                        &                 &                 &  &
	1 &  \beta_{2n}x
	\end{array}
\right), 
\end{equation} 
and we find that the $\alpha_i$'s and $\beta_i$'s are related through
\begin{eqnarray}
\beta_0 &=& \alpha_0, \beta_1 = \frac{1}{\alpha_1}, \ldots,
 \beta_i =\frac{1}{\alpha_i \beta_{i-1}}, \ldots,  \\
\alpha_0 &=& \beta_0,\alpha_1 = \frac{1}{\beta_1}, \ldots, \alpha_i =
\frac{1}{\beta_{i-1} \beta_i }, \ldots. 
\end{eqnarray} 
In order to understand the relation
between different continued fraction representations of the same rational
function in detail we need to take a closer look at the properties of continued
fractions (see e.g.\cite{Wa48,Jo80}).

%%%%%%%%%%%%%%%%%%%%%%%%%%%%%%%%%%%%%%%%%%%%%%%%%%%%%%%%%%%%%%%%%%%%%%%%%
\section{Continued fractions}
\label{sec:continued fractions}
%%%%%%%%%%%%%%%%%%%%%%%%%%%%%%%%%%%%%%%%%%%%%%%%%%%%%%%%%%%%%%%%%%%%%%%%%
A generic (truncated) continued fraction $\frac{A_n}{B_n}$ is
conveniently written as
\begin{equation} 
\frac{A_n}{B_n} = \beta_0 +
\frac{\alpha_1|}{|\beta_1} +  \frac{\alpha_2|}{|\beta_2}
\ldots + \frac{\alpha_n}{\beta_n} \,.
\end{equation}
Simple continued fractions have the property $\alpha_i=1, i=1,..., n$
and one usually writes
\begin{equation} 
\frac{A_n}{B_n} = [\beta_0; \beta_1, \beta_2, \ldots, \beta_n] \,.
\end{equation}

Continued fractions are widely used in many areas of physics and
mathematics, in particular also in number theory. Finite continued
fractions provide an alternative representation of rational numbers
and form the basis of rational approximation theory.  Infinite
continued fractions on the other hand can be used to represent
irrational numbers.  Some numbers have beautiful continued fraction
expansions while others have very mysterious ones. Let us quickly note
a few examples for our amusement:
\begin{eqnarray}
	\phi &=& [1;1,1,1,1,\ldots], \nonumber\\ 
	\sqrt{2} &=& [1;2,2,2,2,\ldots], \nonumber	 \\ 
	   e &=& [2; 1,2,1, 1,4,1, 1,6,1, 1,8,1, 1,10,1,\ldots], \nonumber \\ 
	 \pi &=& [3;7,15,1,292,1,1,1,2,1,3,1,\ldots], \nonumber 
\end{eqnarray} 
where $\phi = \frac{1}{2}(1+\sqrt{5})$ is the golden mean. It is
interesting to note that there is no regular pattern known for
$\pi$, and it is not known why this is so.

Evaluation of continued fractions can be done through forward or
backward recurrence algorithms, and the former makes use of the
intimate relation between continued fractions and the coupled two term
relations
\begin{eqnarray}\label{eq:iterative relation 1}
A_{-1} &=& 1, \, A_0 = \beta_0, \, B_{-1} = 0, \, B_0 = 1, \\
A_i & = & \beta_i A_{i-1} + \alpha_i A_{i-2}, \quad i=1,2,3,\ldots,  \\
B_i & = & \beta_i B_{i-1} + \alpha_i B_{i-2},  \quad i=1,2,3,\ldots,
\end{eqnarray}
which can equivalently be written as an iterative matrix equation,
\begin{equation} 
\left( \begin{array}{c}
A_i \\
B_i
\end{array} \right) = \left( \begin{array}{cc}
A_{i-1} & A_{i-2} \\
B_{i-1} & B_{i-2}
\end{array} \right) \left( \begin{array}{c}
\beta_i \\
\alpha_i
\end{array} \right) .
\end{equation}
There is also an interesting connection between continued fractions,
Moebius transformations and the corresponding unimodular matrices.

The natural arithmetic operation for continued fractions is inversion
and the corresponding rule is particularly simple:
\begin{equation}
{[} \beta_0; \beta_1, \ldots {]}^{-1} = \left\{
\begin{array}{ll}
 {[} 0; \beta_0, \beta_1, \ldots{]} & \quad \text{if} \,
 \beta_0 \neq 0,  \\
 {[} \beta_1; \beta_2, \ldots {]} & \quad \text{if} \,
 \beta_0 = 0.
\end{array} \right. 
\end{equation} 
It is also helpful to write down the rule for the multiplication of a
continued fraction by a constant $c$,
\begin{equation}
c \cdot {[} \beta_0; \beta_1, \ldots, \beta_n {]}  = {[} c \cdot \beta_0;
\frac{\beta_1}{c}, c \cdot \beta_2, \frac{\beta_3}{c}, c \cdot \beta_4, \ldots
{]}.
\end{equation}
Most important for our purpose, however, is the equivalence
transformation of a continued fraction which is stated in the
following theorem:
{\theorem
Two continued fractions $\beta_0+ \frac{\alpha_1|}{|\beta_1}+ \ldots+
\frac{\alpha_n}{\beta_n}$ and $\beta_0' +
\frac{\alpha_1'|}{|\beta_1'}+ \ldots+ \frac{\alpha_n'}{\beta_n'}$ are
equivalent iff there exists a sequence of non--zero constants ${c_n}$
with $c_0 = 1$ such that
\begin{eqnarray} 
\alpha_i'  = & c_i c_{i-1} \alpha_i, & i=1,2,3, \ldots, n, \\
\beta_i'  = & c_i \beta_i, & i=0,1,2, \ldots, n. 
\end{eqnarray}}
The theorem is easily seen to hold true by explicitly writing
out the full continued fraction,
\begin{equation} 
\beta_0 + \cfrac{\alpha_1}{\beta_1 +
\cfrac{\alpha_2}{\beta_2+\cfrac{\cdots}{\cdots +
\cfrac{\alpha_n}{\beta_n}}}}  = \beta_0 + \cfrac{c_1 \alpha_1}{c_1 \beta_1 +
\cfrac{c_1 c_2 \alpha_2}{c_2 \beta_2+\cfrac{\cdots}{\cdots +
\cfrac{c_{n-1} c_n \alpha_n}{c_n \beta_n}}}}  
\end{equation}
and it is also clear that in terms of approximants we simply have
$\frac{A_n}{B_n} =\frac{c_1 c_2 \ldots c_n A_n}{c_1 c_2 \ldots c_n
B_n}$. So we find that a given rational function corresponds to an
equivalence class of continued fractions and the class is parametrised
by the (non--zero) coefficients $c_i$. While the equivalence
transformation itself appears to be rather trivial, and indeed leaves
the value of the continued fraction invariant, it affects the spectrum
of the corresponding matrix representation \cite{Bo02}.

In the analytic theory of continued fractions one considers continued
fractions of the form
\begin{equation}
\beta_0(z) + \frac{\alpha_1(z)|}{|\beta_1(z)} +
\frac{\alpha_2(z)|}{|\beta_2(z)} + \ldots ,
\end{equation} 
i.e., the coefficients are functions of a complex variable
$z$. So--called J--fractions are of the special form
\begin{equation} 
{[}r_1 z + s_1, r_2 z + s_2, r_3 z + s_3, \ldots{]}, 
\end{equation} 
where $r_i,s_i$ are complex numbers with $r_i \neq 0$. One can show
that the $n$-th approximant of a J--fraction is an element of
$R_{n-1,n}$.  Conversely, let
\begin{eqnarray}
P_{n-1}(z) & = & p_1 z^{n-1} + p_2 z^{n-2} + \ldots + p_n \, ,\\
Q_n(z) & = & q_0 z^n + q_1 z^{n-1} + \ldots + q_n \, , 
\end{eqnarray}
then we have
\begin{equation}
\frac{P_{n-1}(z)}{Q_n(z)} = {[}r_1 z + s_1, r_2 z + s_2, \ldots, 
       r_n z + s_n {]}
\end{equation}
where $r_i, s_i$ are uniquely determined by $p_i,q_i$.  We can now
replace $z$ by $-z$, apply an appropriate equivalence transformation
and, using the fact that $P_{n-1}(z)/Q_n(z)$ is odd, we find $s_i = 0$
from uniqueness. Therefore we can always write
\begin{equation}
 z \frac{P_{n-1}(z^2)}{Q_n(z^2)} = {[}k_1 z, k_2 z, k_3 z, \ldots, k_m
z {]} \, ,
\end{equation}
where $m=2n-1$ or $m=2n$ according as $Q_n(0)=0$ or $Q_n(0) \neq 0$.

%%%%%%%%%%%%%%%%%%%%%%%%%%%%%%%%%%%%%%%%%%%%%%%%%%%%%%%%%%%%%%%%%%%%%%%%%
\section{Application to the overlap operator}
%%%%%%%%%%%%%%%%%%%%%%%%%%%%%%%%%%%%%%%%%%%%%%%%%%%%%%%%%%%%%%%%%%%%%%%%%
We are now in a position to apply our knowledge to find the solution
to the equation $\frac{2}{1-\mu} \gamma_5 D(\mu) \psi =
\chi$. Collecting the results from the previous two sections we obtain
an equivalence class of five dimensional linear systems
\begin{equation} \nonumber
{\small
\left(  \begin{array}{cccccc}
  A  \gamma_5 + k_0 H & c_1 &          & & & \\
  c_1                 & - c_1^2 k_1 H  &  c_1 c_2 & & & \\
                                 &   c_1 c_2      & c_2^2 k_2 H &  & & \\ 
                                 &                &             &  \ddots & & \\
  & & & & - c_{2n-1}^2 k_{2n-1} H & c_{2n-1} c_{2n} \\
  & & & & c_{2n-1} c_{2n} & c_{2n}^2 k_{2n} H \\
	\end{array} \right) \left(\begin{array}{c} 
					\psi \\
					\phi_1 \\
                                        \phi_2 \\
                                        \vdots \\
					\phi_{2n-1} \\
					\phi_{2n} 
				\end{array}\right) 
=
\left(\begin{array}{c} 
					\chi \\
					0    \\
                                        0    \\
	                                \vdots\\
                                        0    \\
	                                0
				\end{array}\right)   
}
\end{equation}
where the $k_i$'s and $n$ are uniquely determined by the given
rational approximation to the sign--function, the $c_i$'s parametrise
the corresponding equivalence class and $A=\frac{1+\mu}{1-\mu}$. It is
now crucial to see how the spectrum of the five dimensional matrix
depends on the free parameters $c_i$ as we already emphasised in the
last section. While for a generic set of parameters the five
dimensional system is usually ill--defined, the condition number can
be kept under control with a clever choice of $c_i$'s \cite{Bo02,We04}
enabling one to optimise the matrix, e.g., for fast inversions. As was
pointed out in \cite{Bo02} the equivalence transformations can be
understood as a block Jacobi preconditioning without any computational
overhead. The particularly simple structure of the five dimensional
operator allows improvements in various directions: one can easily
change and optimise the hermitian overlap kernel $H$ or apply various
well known preconditioning techniques such as an even--odd or ILU
decomposition \cite{We04}.

It is instructive to see that the first auxiliary field disentangles
$\gamma_5$ from the sign--function, i.e., $(A \cdot \gamma_5 +
\text{sign}(H)) \psi = \chi$ maps into
\begin{equation}
\left( \begin{array}{cc} A \cdot \gamma_5 & c \\ c & - c^2
	\text{sign}(H) \end{array} \right) \left(\begin{array}{c} \psi
	\\ \phi \end{array}\right) = \left(\begin{array}{c} \chi \\ 0
	\end{array}\right)
\end{equation}
where we used $\text{sign} = \text{sign}^{-1}$. Additional fields are
then only used to generate the $\text{sign}$--function (or its
approximation, respectively). We now have two systems which can be
easily solved,
\begin{equation} 
\psi = \frac{1}{A} \gamma_5 (\chi - c \phi), \quad
\phi = \frac{1}{c} \text{sign}(H) \psi,
\end{equation}
yielding recursion relations for $\psi$ and $\phi$,
\begin{equation} 
\psi^{(i+1)} = \frac{1}{A} \gamma_5 ( \chi - \text{sign}(H) \psi^{(i)}), \quad
\phi^{(i+1)} = \frac{1}{A} \text{sign}(H) \gamma_5 ( \frac{1}{c} \chi - \phi^{(i)}).
\end{equation}
Equivalently, there are recursion relations for the residuals and one
can show that
\begin{equation}
|r_{\psi,\phi}^{(i)}| = \Big(\frac{1-\mu}{1+\mu} \Big)^i |r_{\psi,\phi}^{(0)}|.
\end{equation}
Of course one can use a similar recursive scheme for the case where
$\text{sign}(H)$ is expressed as a continued fraction and one has
several auxiliary fields.

Projection of eigenvectors of $H$ close to 0 is a valuable tool to
improve approximations to the overlap operator and here it is
straightforward to implement. However, we wish to point out that in
this formulation it might not be necessary at all. Consider the linear
system in the lower right corner,
\begin{equation}
c_{2n-1} c_{2n} \phi_{2n-1} -
c_{2n}^2  k_{2n} H \phi_{2n} = 0.
\label{eq:last_system}
\end{equation}
For a typical rational approximation to the sign--function we have
$k_{2n} \gg 1$ and with the choice $c_{2n} \simeq 1/\sqrt{k_{2n+1}}
\ll 1$. So it turns out that the system in eq.(\ref{eq:last_system})
is essentially equivalent to finding eigenvectors of $H$ close to
zero, i.e., the two Krylov spaces possibly have a large
overlap. Indeed a truncation in the fifth dimension, i.e., of a given
continued fraction, changes the approximation only in the
neighbourhood around $H=0$ and therefore provides a natural truncation
scheme for approximations of fixed accuracy. It also opens up the
possibility of applying successively better approximations to the
overlap operator which are ultra--local in five dimensions.

%%%%%%%%%%%%%%%%%%%%%%%%%%%%%%%%%%%%%%%%%%%%%%%%%%%%%%%%%%%%%%%%%%%%%%%%%
\section{Summary and outlook}
%%%%%%%%%%%%%%%%%%%%%%%%%%%%%%%%%%%%%%%%%%%%%%%%%%%%%%%%%%%%%%%%%%%%%%%%%
We have shown how to use a continued fraction expansion of the
sign--function in order to obtain a five dimensional formulation of
the overlap lattice Dirac operator. We have pointed out that the
operator can be made well conditioned using equivalence
transformations on the continued fractions. It is now important to
investigate in detail strategies to exploit this freedom in practical
applications and such a study is under way \cite{We04}. If successful,
and first results indicate that this is indeed the case, the
formulation would provide a valuable alternative for the simulation of
dynamical chiral fermions.

%%%%%%%%%%%%%%%%%%%%%%%%%%%%%%%%%%%%%%%%%%%%%%%%%%%%%%%%%%%%%%%%%%%%%%%%%
\section[]{Acknowledgments}
%%%%%%%%%%%%%%%%%%%%%%%%%%%%%%%%%%%%%%%%%%%%%%%%%%%%%%%%%%%%%%%%%%%%%%%%%
I would like to thank the organisers of the workshop for the
invitation and for creating a stimulating atmosphere. I also thank
Tony Kennedy for discussions and comments. This research was supported
by a PPARC SPG fellowship.

% BibTeX users please use
% \bibliographystyle{}
% \bibliography{}
%
% Non-BibTeX users please follow the syntax
% the syntax of "referenc.tex" for your own citations

%%%%%%%%%%%%%%%%%%%%%%%%%%%%%%%%%%%%%%%%%%%%%%%%%%%%%%%%%%%%%%%%%%%%%%

%%%%%%%%%%%%%%%%%%%%%%%%%%%%%%%%%%%%%%%%%%%%%%%%%%%%%%%%%%%%%%%%%%%%%%

%\printindex
\end{document}